\documentclass[epj,final]{svjour}
\usepackage{latexsym}
\usepackage{amsmath}
\usepackage{amssymb}
\usepackage{graphics}

\begin{document}

\title{Spontaneous distortion in the spin-1/2 Ising-Heisenberg model on 
decorated planar lattices with a magnetoelastic coupling}
\author{Jozef Stre\v{c}ka\inst{1} \fnmsep \thanks{\email{jozef.strecka@upjs.sk}} 
        \and Onofre Rojas\inst{2} \and S.M. de Souza\inst{2}}                     
\institute{Department of Theoretical Physics and Astrophysics, Faculty of Science, \\ 
P. J. \v{S}af\'{a}rik University, Park Angelinum 9, 040 01 Ko\v{s}ice, Slovak Republic \and 
Departamento de Ciencias Exatas, Universidade Federal de Lavras, CP 3037, 37200000, MG, Brazil}
\authorrunning{\textit{J. Stre\v{c}ka} et al.}
\titlerunning{Spontaneous distortion in the spin-1/2 Ising-Heisenberg model with a magnetoelastic coupling}
\date{Received: date / Revised version: date}
\abstract{Magnetoelastic properties of the spin-1/2 Ising-Heisenberg model on doubly decorated planar lattices partially amenable 
to lattice vibrations are examined within the framework of the harmonic approximation and decoration-iteration transformation. 
It is shown that the magnetoelastic coupling may lead to a spontaneous distortion of the vibrating decorating atoms and the mutual 
interplay between quantum spin fluctuations and local lattice deformations enhances typical quantum features like the quantum reduction 
of the magnetization in the ground state of the quantum antiferromagnetic phase, while it does not affect the ground-state behaviour of 
the classical ferromagnetic phase. It also turns out that the spontaneous distortion is responsible for a much more pronounced 
reduction of the critical temperature in the quantum antiferromagnetic phase than in the classical ferromagnetic phase.} 
%\keywords{Ising-Heisenberg model--magnetoelastic coupling--critical phenomena}

\maketitle

\section{Introduction}
\label{intro}

Classical and quantum spin models traditionally attract an appreciable scientific interest as they 
bring an insight into various aspects of phase transitions, critical phenomena and many other diverse 
phenomena of a cooperative nature \cite{matt93,lavi99,diep04,wu09}. However, the predominant number 
of suggested lattice-statistical spin models tacitly assumes that atomic spins are placed at rigid lattice positions and they interact among themselves by means of a constant (usually nearest-neighbour) exchange interaction. This conjecture represents an obvious idealization, because atoms are carrying out persistent oscillatory motion around their equilibrium lattice positions even at zero temperature and the exchange interaction may strongly depend on the interatomic distance \cite{ohan00}. Early attempts to deal with magnetoelastic properties of compressible spin models were based on a rather simple but unrealistic 
assumption that the exchange interaction between the atomic spins depends just on the average volume \cite{rice54,domb56,bean62}. To take into account the effect of fluctuations of lattice spacings on a magnitude of the exchange interaction, the more realistic compressible spin models were later proposed 
on assumption that the exchange interaction varies linearly with an instantaneous interatomic distance \cite{matt63,bake70,gunt71,mats68,sali74,pens73}. Among other matters, it has been shown that the 
spin-lattice interaction may change a continuous (second-order) phase transition to a discontinuous (first-order) one \cite{imry74,berg76}, provoke an interesting renormalization of critical exponents 
\cite{pens73,fish68}, or cause the lattice instability with respect to a spontaneous spin-Peierls 
dimerization \cite{pinc71,beni72,tayl85,okam92,derz00}.

It is worthy to notice that almost all exactly tractable spin models including the spin-lattice interaction are of a classical nature such as the Ising models \cite{matt63,bake70,gunt71,enti73,sali73} or the classical Heisenberg models \cite{lee73,barm75,pens76,marc85,kapl09}. Despite a considerable effort, rigorous treatment of quantum spin models with the spin-lattice interaction is usually incredible task except of a few highly valuable cases like the spin-1/2 XX chain in a transverse magnetic field displaying the spin-Peierls phenomenon \cite{pinc71,beni72,tayl85,okam92,derz00} or the compressible version of the spin-1/2 XX chain 
in a transverse magnetic field \cite{orig05}. Accordingly, one usually has to rely on application 
of some approximative method when treating a quantum spin model with the spin-lattice interaction \cite{pytt65,pytt74,beni73,cros79,uhri98,augi98,onis03,akiy11}. 

An important challenge for theoretical studies currently represents an investigation of the effect 
of the spin-lattice interaction on ground state, phase transitions and critical phenomena of quantum spin models. 
Bearing this in mind, it is also highly desirable to search for simpler exactly tractable quantum 
spin models with the spin-lattice interaction, which could provide a deeper understanding of the effect 
of a mutual interplay between quantum spin fluctuations and fluctuations of lattice spacings on a critical behaviour. 
The hybrid Ising-Heisenberg models defined on planar lattices, which are occupied in part by the classical Ising spins and partly by the quantum Heisenberg spins, belong to a few eligible candidates capable of this study \cite{stre02,stre06,stre08,yao08,valv09,stre09,sun11}. It is noteworthy that the exactly solved Ising-Heisenberg planar models have already proved their usefulness in predicting several unusual ordered
or disordered quantum phases \cite{stre02,stre06,stre08,yao08}, explaining a partial lifting of the ground-state degeneracy in disordered spin-liquid phases \cite{stre08,yao08}, providing a deeper insight into a weak-universal critical behaviour \cite{valv09,stre09} or quantum correlations 
from the viewpoint of quantum computation science \cite{sun11}. The main objective of the present work 
is to extend the spin-1/2 Ising-Heisenberg model on doubly decorated planar lattices from Ref. \cite{stre02} by accounting for the spin-lattice interaction treated within the harmonic approximation. Our calculation will closely follow the approach invented by Bellucci and Ohanyan \cite{bell10} for the spin-1/2 Ising-Heisenberg sawtooth chain with the spin-lattice interaction except that the spin degrees of freedom 
of the Heisenberg spins will be treated before integrating out the lattice degrees of freedom.\footnote{Note that the reverse procedure as used in Ref. \cite{bell10} is not actually exact, because one cannot simply factorize the bond partition function into two parts (one of which is independent of the distortion parameter, while the other one does depend on the distortion parameter) before integrating out the lattice degrees of freedom due to a non-commutability of both parts (perturbed and unperturbed) of the same bond Hamiltonian.} 

Finally, it is worthwhile to remark that the magnetic structure of the model under investigation has been inspired by two layered metal frameworks of polymeric coordination compounds 2Rh$_2$(CH$_3$COO)$_4$ $\cdot$ K$_3$Co(CN)$_6$ \cite{lu96} and [Cu$_2$(L)]$_3$[Co(CN)$_6$]$_2$ $\cdot$ 6CH$_3$OH $\cdot$ 2DMF \cite{miya05} in which metal ions 
are arranged so as to build the doubly decorated square lattice and the doubly decorated honeycomb lattice, respectively. Unfortunately, the trivalent Co$^{3+}$ ions are 
diamagnetic in both aforementioned compounds due to a strong ligand field of the cyano group and hence, there does not exist nowadays any appropriate results 
of magnetic measurements to compare with our theoretical predictions. However, the recent progress in a targeted design of molecular-based magnetic materials with desired crystal structure and magnetic properties gives us hope for a preparation of experimental representative(s) of the model under investigation \cite{verd01,cern02,lesc05}. 
As a matter of fact, the rational synthesis of isostructural analogs of both aforementioned polymeric complexes \cite{lu96,miya05} that would contain within the complex anion [M(CN)$_6$]$^{3-}$ the low-spin Fe$^{3+}$ ion ($S$=1/2) instead of the diamagnetic Co$^{3+}$ ion ($S$=0), represents one of the most promissing routes for a design of suitable model compound and might be regarded as a challenge for molecular engineering. Besides, it is also our hope that the results presented in the following may bring insight into many important features arising from the spin-lattice interaction, which could manifest themselves in a wider class of insulating magnetic materials. 

The outline of this paper is as follows. In Section \ref{sec:model}, we first provide a rather 
detailed description of the model under investigation together with the most crucial steps of the 
calculation procedure, which enables us to explore ground-state and finite-temperature phase diagrams, 
the relevant behaviour of order parameters and the distortion parameter. The most interesting results 
are then presented and detailed discussed in Section \ref{sec:result}. Finally, the summary of 
the most important scientific achievements is mentioned with several concluding remarks in Section \ref{sec:conc}.

\section{Model and its solution}
\label{sec:model}

\begin{figure}
\begin{center}
\resizebox{0.7\columnwidth}{!}{\includegraphics{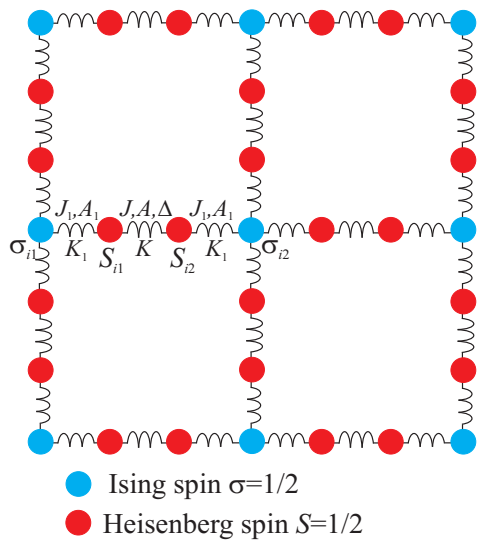}}
\end{center}
\vspace{-0.3cm}
\caption{The segment from the doubly decorated square lattice. Blue circles label rigid lattice positions of the Ising spins $\sigma=1/2$, while the red ones denote lattice positions of the decorating Heisenberg spins $S=1/2$ prone to lattice vibrations.}
\label{fig:1}       
\end{figure}

Consider the spin-1/2 Ising-Heisenberg model defined on doubly decorated planar lattices as schematically illustrated in Fig.~\ref{fig:1} on a particular example of the doubly decorated square lattice. Within the framework of this model, each nodal site of some doubly decorated planar lattice is occupied by the Ising atom with the spin $\sigma = 1/2$ while each its bond incorporates a couple of the quantum Heisenberg atoms with the spin $S=1/2$. Suppose furthermore that the Ising atoms are placed at rigid lattice positions in contrast with the decorating Heisenberg atoms, which are relaxed from the condition of a perfect lattice rigidity and may consequently change their equilibrium lattice positions after taking into account the elastic energy penalty for each distorted bond within the harmonic approximation. It is noteworthy that the assumption of lattice rigidity of the Ising atoms perhaps represents the most important drawback of our model, which is however quite reasonable if: (i) the mass of the Ising atoms is much larger than the mass of the Heisenberg atoms; (ii) the model is defined on doubly decorated planar lattices with a high enough coordination number of the Ising atoms (the displacement of the Ising atoms from their equilibrium lattice positions would then cost much larger amount of the elastic energy owing to a deformation of greater number of bonds); (iii) the spring stiffness constant of the Ising bonds is much stronger than the elastic constant of the Heisenberg bonds; 
(iv) various combinations of previous three particular cases. 

For further convenience, let us write the total Hamiltonian of the investigated model system as a sum of bond Hamiltonians 
\begin{equation}
{\cal H} = \sum_{i=1}^{Nq/2} {\cal H}_i = \sum_{i=1}^{Nq/2} ({\cal H}_i^m + {\cal H}_i^e).
\label{ha}
\end{equation}
Each bond Hamiltonian ${\cal H}_i$, which is further splitted into the magnetoelastic part ${\cal H}_i^m$ and the pure elastic part ${\cal H}_i^e$, contains all the interaction terms of one couple of the Heisenberg spins from the $i$th bond (see Fig. \ref{fig:1}) [$N$ denotes the total number of the Ising spins and $q$ determines the number of their nearest neighbours]. The magnetoelastic part of the bond Hamiltonian is given by
\begin{eqnarray}
{\cal H}_i^m = &-& [J + A (x_{i1} + x_{i2})] 
  [\Delta ({S}_{i1}^x {S}_{i2}^x + {S}_{i1}^y {S}_{i2}^y ) 
                  + {S}_{i1}^z {S}_{i2}^z ] \nonumber \\
                  &-& (J_1 - A_1 x_{i1}) {S}_{i1}^z {\sigma}_{i1}^z
                  - (J_1 - A_1 x_{i2}) {S}_{i2}^z {\sigma}_{i2}^z.                  
\label{ham}
\end{eqnarray}
Here, ${S}_{i\alpha}^{\gamma}$ and ${\sigma}_{i\alpha}^{z}$ ($\alpha=1,2$, $\gamma=x,y,z$)
denote spatial components of the standard spin-1/2 operator, the parameter $J$ labels XXZ interaction 
between the nearest-neighbour Heisenberg spins, $\Delta$ is a spatial anisotropy in this interaction 
and the parameter $J_1$ stands for the Ising interaction between the nearest-neighbour Ising and Heise\-nberg spins. Furthermore, the parameter $x_{i\alpha}$ denotes a displacement of the Heisenberg spin $S_{i\alpha}$ from its equilibrium position referred relative to its nearest-neighbor Ising spin $\sigma_{i\alpha}$, 
whereas $x_{i\alpha}>0$ ($x_{i\alpha}<0$) corresponds to an elongation (contraction) of the relevant distance. 
It is worthwhile to remark that only a rather small lattice displacements are feasible  within the harmonic approximation, which consequently allows one to perform a linear expansion of the distant-dependent 
exchange interactions around their non-distorted values $J$ and $J_1$. Hence, it follows that a linear increase (decrease) in the Heisenberg and Ising interaction originates from a contraction (elongation) 
of a relevant bond length and this distortion-induced change in the exchange interactions $J$ and $J_1$ 
is proportional to the magnetoelastic (spin-lattice) couplings $A$ and $A_1$, respectively. Finally, 
the pure elastic part of the bond Hamiltonian takes into account a kinetic energy of both Heisenberg 
atoms prone to lattice oscillations as well as the elastic energy penalty for one Heisenberg bond and 
two Ising bonds to be treated as classical harmonic oscillators 
\begin{eqnarray}
{\cal H}_i^e = \frac{p_{i1}^2 + p_{i2}^2}{2M} + \frac{K}{2}  \left(x_{i1} + x_{i2}\right)^2 
                 + \frac{K_1}{2}  \left(x_{i1}^2 + x_{i2}^2\right).                  
\label{hae}
\end{eqnarray}
Above, $p_{i\alpha} = M \frac{{\rm d} x_{i\alpha}}{{\rm d} t}$ denotes the momentum of the Heisenberg atoms with the mass $M$, $K$ and $K_1$ label spring stiffness (bare elastic constants) of the Heisenberg and Ising bond, respectively. The first term thus represents the kinetic energy of both the Heisenberg atoms, the second term stands for the elastic energy of the Heisenberg bond and the last third term represents the elastic energy of both Ising bonds. Last but not least, it is worthy to note that the elastic part of the bond Hamiltonian (\ref{hae}) takes into consideration only 'longitudinal' lattice deformations in directions of lattice bonds, which should provide the most important effect because in-plane or out-of-plane 'transverse' lattice deformations would couple quadratically in the Hamiltonian and thus, these less important higher-order terms would just effectively cause rescalling of the elastic term only.  

The crucial step of our procedure represents calculation of the partition function. Different bond Hamiltonians obviously commute with each other, i.e. $[{\cal H}_i, {\cal H}_j] = 0$, and in the consequence of that, the total partition function can be calculated when summing 
the product of bond partition functions over all possible spin configurations of the Ising spins 
\begin{eqnarray}
{\cal Z} = \sum_{\{ \sigma_i \}} \prod_{i=1}^{Nq/2} {\cal Z}_i.       
\label{z}
\end{eqnarray}
Each bond partition function ${\cal Z}_i$ in fact represent the local partition function 
of one couple of the Heisenberg spins from the $i$th bond and can be defined as follows
\begin{eqnarray}
{\cal Z}_i = \!\! \iint\limits_{\!\!\!\!\!-\infty}^{\phantom{100}\infty} \!\!
  \frac{{\rm d} x_{i1} {\rm d} p_{i1}}{h} \!\! \iint\limits_{\!\!\!\!\!-\infty}^{\phantom{100}\infty} \!\! 
\frac{{\rm d} x_{i2} {\rm d} p_{i2}}{h} \exp(-\beta {\cal H}_i^e) 
{\rm Tr}_{i} \exp(-\beta {\cal H}_i^m).      \nonumber \\ 
\label{zi}
\end{eqnarray}
Here, $\beta = 1/(k_{\rm B} T)$, $k_{\rm B}$ is Boltzmann's constant, $T$ is the absolute temperature, $h$ is Planck's constant and the symbol ${\rm Tr}_{i}$ means a trace over spin degrees of freedom of the $i$th couple of the Heisenberg spins. 

To proceed further with calculations, it is advisable to perform local canonical coordinate 
transformation unambiguously given by
\begin{eqnarray}
x_{i1} &=& \frac{1}{\sqrt{2}} \left(y_{i1} + y_{i2} \right), \qquad
x_{i2} = \frac{1}{\sqrt{2}} \left(y_{i1} - y_{i2} \right);       
\label{cordtrans1}
\end{eqnarray}
or its inverse
\begin{eqnarray}
y_{i1} &=& \frac{1}{\sqrt{2}} \left(x_{i1} + x_{i2} \right), \qquad
y_{i2} = \frac{1}{\sqrt{2}} \left(x_{i1} - x_{i2} \right);        
\label{cordtrans2}
\end{eqnarray}
which introduces two new spatial coordinates $y_{i1}$ and $y_{i2}$ instead of local displacement parameters $x_{i1}$ and $x_{i2}$. The spatial coordinate $y_{i1}$ is connected to an average displacement 
of both Heisenberg spins from their equilibrium lattice positions, while the other spatial coordinate 
$y_{i2}$ reflects a possible assymetry in the displacement of both Heisenberg spins relative with 
respect to their nearest Ising neighbors. The main advantage of the canonical coordinate transformation (\ref{cordtrans1}) lies in a separable form of the potential energy from the elastic part 
of the bond Hamiltonian (which subsequently allows an independent integration over both new spatial coordinates $y_{i1}$ and $y_{i2}$) while keeping the kinetic term unchanged. Applying the canonical 
coordinate transformation (\ref{cordtrans1}), the elastic part of the bond Hamiltonian (\ref{hae}) 
can be actually rewritten using the new coordinates $y_{i\alpha}$ and their conjugated momenta
$q_{i\alpha} = M \frac{{\rm d} y_{i\alpha}}{{\rm d} t}$ into the following form
\begin{eqnarray}
{\cal H}_i^e = \frac{q_{i1}^2 + q_{i2}^2}{2M} + K y_{i1}^2 
             + \frac{K_1}{2} \left(y_{i1}^2 + y_{i2}^2\right).                  
\label{haey}
\end{eqnarray}
A substitution of the elastic part of the bond Hamiltonian (\ref{haey}) into Eq.~(\ref{zi}) yields the following result for the bond partition function after a straightforward integration over both momenta  
\begin{eqnarray}
{\cal Z}_i = \frac{2 \pi M}{\beta h^2} \!\!
\int\limits_{-\infty}^{\infty} \!\!\!&&\!\!\! {\rm d} y_{i1}  
\exp \left[- \frac{\beta}{2}  (K_1 + 2K) y_{i1}^2 \right] \nonumber \\
\times \!\! \int\limits_{-\infty}^{\infty} \!\!\!&&\!\!\! 
{\rm d} y_{i2} \exp \left(- \frac{\beta}{2}  K_1 y_{i2}^2 \right) 
{\rm Tr}_{i} \exp(-\beta {\cal H}_i^m).      
\label{ziy}
\end{eqnarray}
It is quite evident from Eq.~(\ref{ziy}) that it is now necessary to find the partial trace over spin degrees of freedom of the $i$th couple of the Heisenberg spins in order to get expression suitable for a subsequent integration over spatial coordinates. This can be rather easily accomplished by diagonalizing the magnetoelastic part of the bond Hamiltonian (\ref{ham}) and considering a trace invariance, which enables 
one to express this partial trace in terms of four eigenvalues of the Hamiltonian ${\cal H}_i^m$
\begin{eqnarray}
{\rm Tr}_{i} \exp(-\beta {\cal H}_i^m) = \sum_{j=1}^{4} \exp(-\beta E_{ij}),       
\label{zim}
\end{eqnarray}
which are explicitly given by the following formulas
\begin{eqnarray}
E_{i1,i2} &=& - \frac{{\cal J}}{4} \pm \frac{{\cal J}_1}{2} \sigma_{i1}^z 
                        \pm \frac{{\cal J}_2}{2} \sigma_{i2}^z,    \label{es1} \\
E_{i3,i4} &=& \frac{{\cal J}}{4} \pm \frac{1}{2} \sqrt{\left( {\cal J}_1 \sigma_{i1}^z 
                 - {\cal J}_2 \sigma_{i2}^z \right)^2 + \left( {\cal J} \Delta \right)^2}.                     \label{es2}
\end{eqnarray}
For the sake of brevity, the four energy eigenvalues (\ref{es1})--(\ref{es2}) are expressed 
in terms of the distant-dependent exchange interactions
\begin{eqnarray}
{\cal J}   &=& J + A (x_{i1} + x_{i2}) = J + \sqrt{2} A y_{i1}, \nonumber \\ 
{\cal J}_1 &=& J_1 - A_1 x_{i1} = J_1 - \frac{A_1}{\sqrt{2}} \left(y_{i1} + y_{i2} \right),    \nonumber \\
{\cal J}_2 &=& J_1 - A_1 x_{i2} = J_1 - \frac{A_1}{\sqrt{2}} \left(y_{i1} - y_{i2} \right).                  
\label{ji}
\end{eqnarray}
It should be noted here that the former couple of eigenvalues (\ref{es1}) varies linearly with the displacement parameters $y_{i1}$ and $y_{i2}$ in contrast to the latter couple of eigenvalues (\ref{es2}) involving both these displacements inside a square root, which would consequently preclude an integration 
over lattice degrees of freedom at the level of the bond partition function. However, the latter two eigenvalues (\ref{es2}) can still be expanded into a power series in terms 
of both displacement parameters $y_{i1}$ and $y_{i2}$, whereas it is sufficient to retain only the 
terms up to the first order and neglect all higher-order terms on behalf of small magnitude of both displacements. Using this approach, one easily obtains four eigenvalues of the Hamiltonian 
${\cal H}_i^m$ that linearly depend on the displacement parameters $y_{i1}$ and $y_{i2}$ 
according to the relations
\begin{eqnarray}
E_{i1,i2} &=& - \frac{J}{4} \mp \frac{J_1}{2} (\sigma_{i1}^z + \sigma_{i1}^z)
                       \pm \frac{y_{k2}}{2 \sqrt{2}} A_1 (\sigma_{i1}^z - \sigma_{i1}^z) \nonumber \\
          &-& \frac{y_{k1}}{2 \sqrt{2}} [A \mp A_1 (\sigma_{i1}^z + \sigma_{i1}^z)],  \label{ees1}    \\       
E_{i3,i4} &=& \frac{J}{4} \mp \frac{1}{2} 
\sqrt{J_1^2 \left(\sigma_{i1}^z - \sigma_{i2}^z \right)^2 + \left(J \Delta \right)^2}  \nonumber \\
&+& \frac{y_{k1}}{2 \sqrt{2}} \left[A \pm \frac{J_1 A_1 (\sigma_{i1}^z + \sigma_{i1}^z)^2 - 2 J A \Delta^2}
{\sqrt{J_1^2 \left(\sigma_{i1}^z - \sigma_{i2}^z \right)^2 + \left(J \Delta \right)^2}}\right]\!\!.          
\label{ees2}
\end{eqnarray} 
At this stage, the four energy eigenvalues given by Eqs.~(\ref{ees1}) and (\ref{ees2}) can be used for calculating the partial trace in the magnetoelastic part of the bond partition function (\ref{zim}) 
and the resultant expression can be further substituted to the bond partition function (\ref{ziy}) 
in order to perform an integration over lattice degrees of freedom. After a straightforward but little bit cumbersome integration over the spatial coordinates $y_{i1}$ and $y_{i2}$ one finally gets the bond partition function 
\begin{eqnarray}
\!\!\!\!&&\!\!\!\! {\cal Z}_i = \frac{2 \exp \left[\frac{\beta A^2}{16(K_1 + 2K)} \right]}{\beta^2 \hbar^2 \omega_1 \omega_2}  \Biggl \{ \exp \Biggl[\frac{\beta J}{4} + \frac{\beta A_1^2 (\sigma_{i1}^z + \sigma_{i2}^z)^2}{16(K_1+2K)} +
\nonumber \\
\!\!\!\!&&\!\!\!\! \frac{\beta A_1^2 (\sigma_{i1}^z - \sigma_{i2}^z)^2}{16K_1} \Biggr]   
\cosh \left[ \frac{\beta}{8} \left(4 J_1 - \frac{A A_1}{K_1 + 2K} \right) 
(\sigma_{i1}^z + \sigma_{i2}^z) \right] \nonumber \\
\!\!\!\!&&\!\!\!\! + \exp\left[-\frac{\beta J}{4} + \frac{\beta [J_1 A_1 (\sigma_{i1}^z - \sigma_{i2}^z)^2 - 2 J A \Delta^2]^2} {16(K_1+2K) [J_1^2 (\sigma_{i1}^z - \sigma_{i2}^z)^2 + (J \Delta)^2] }\right] 
\nonumber \\
\!\!\!\!&&\!\!\!\! \times \cosh \Biggl[ \frac{\beta}{2} \sqrt{J_1^2 \left(\sigma_{i1}^z - \sigma_{i2}^z \right)^2 + \left(J \Delta \right)^2}  \nonumber \\
\!\!\!\!&&\!\!\!\! + \frac{\beta A}{8(K_1 + 2K)} \frac{J_1 A_1 (\sigma_{i1}^z - \sigma_{i2}^z)^2 - 2 J A \Delta^2}{\sqrt{J_1^2 \left(\sigma_{i1}^z - \sigma_{i2}^z \right)^2 + \left(J \Delta \right)^2}}  \Biggr]
\Biggr\},
\label{bpf}
\end{eqnarray} 
whereas $\omega_1 = \sqrt{K_1/M}$ and $\omega_2 = \sqrt{(K_1+2K)/M}$ denote two angular frequencies that correspond to out-of-phase and in-phase normal mode vibrations of both Heisenberg spins. 
It is of fundamental importance that the bond partition function (\ref{bpf}) does not already depend neither on the Heisenberg spin degrees of freedom nor the lattice degrees of freedom, but only upon the two Ising spins $\sigma_{i1}$ and $\sigma_{i2}$. Accordingly, one may exploit the generalized decoration-iteration transformation \cite{fish59,roja09,stre10,lap} in order to substitute the bond partition function (\ref{bpf}) by a simpler equivalent expression depending just on two Ising spins $\sigma_{i1}$ and $\sigma_{i2}$ 
\begin{eqnarray}
{\cal Z}_i (\sigma_{i1}, \sigma_{i2}) = R_0 \exp(\beta R_1 \sigma_{i1} \sigma_{i2}).       
\label{dit}
\end{eqnarray} 
The physical meaning of the generalized decoration-itera\-tion transformation (\ref{dit}) lies in replacing 
all the interaction parameters belonging to the $i$th bond Hamiltonian ${\cal H}_i$ through the unique effective interaction $R_1$ between the two nodal Ising spins $\sigma_{i1}$ and $\sigma_{i2}$. This mapping transformation should of course hold independently of spin states of both these Ising spins and this 'self-consistency' condition \cite{fish59,roja09,stre10,lap} unambiguously determines yet unknown mapping parameters $R_0$ and $R_1$ 
\begin{eqnarray}
R_0 &=& 2 \exp \left[\frac{\beta A^2}{16(K_1 + 2K)} \right] 
\frac{(V_1 V_2)^{\frac{1}{2}}}{\beta^2 \hbar^2 \omega_1 \omega_2}, \label{mpr1} \\ 
\beta R_1 &=& 2 \ln\left(\frac{V_1}{V_2}\right),       
\label{mpr2}
\end{eqnarray}
which are expressed by means of the functions $V_1$ and $V_2$  
\begin{eqnarray}
\!\!\!&&\!\!\! V_1 = \exp \! \left[\frac{\beta J}{4} + \frac{\beta A_1^2}{16 (K_1 + 2 K)} \right] 
\! \cosh \! \left[ \frac{\beta J_1}{2} - \frac{\beta A A_1}{8(K_1 + 2K)}\right] \nonumber \\ \!\!\!&&\!\!\! 
+ \exp \! \left[- \frac{\beta J}{4} + \frac{\beta A^2 \Delta^2}{4 (K_1 + 2 K)} \right] 
\! \cosh \! \left[ \frac{\beta J \Delta}{2} - \frac{\beta A^2 \Delta}{4(K_1 + 2K)}\right]\!, \nonumber \\
\!\!\!&&\!\!\! V_2 = \exp \! \left[\frac{\beta J}{4} + \frac{\beta A_1^2}{16K_1} \right] 
+ \exp \! \left[-\frac{\beta J}{4} + \frac{\beta P_1^2}{16(K_1 + 2 K)} \right] \nonumber \\ \!\!\!&&\!\!\! 
\times \cosh \! \left[ \frac{\beta}{2} \sqrt{J_1^2 + (J \Delta)^2} + \frac{\beta A P_1}{8(K_1 + 2K)} \right]\!, \label{v1v2}
\end{eqnarray}
and the auxiliary function $P_1$ defined as
\begin{equation}
P_1 = \frac{J_1 A_1 - 2 J A \Delta^2}{\sqrt{J_1^2 + (J \Delta)^2}}. 
\label{p}
\end{equation} 

If one substitutes the decoration-iteration transformation (\ref{dit}) satisfying the 'self-consistency' condition (\ref{mpr1})--(\ref{p}) into the partition function (\ref{z}), one readily gains a mapping relationship between the partition function ${\cal Z}$ of the spin-1/2 Ising-Heisenberg model on 
the doubly decorated planar lattices amenable to lattice vibrations and respectively, the partition function ${\cal Z}_{\rm IM}$ of the equivalent spin-1/2 Ising model on a corresponding undecorated 
and rigid planar lattice
\begin{eqnarray}
{\cal Z} (\beta, J, J_1, \Delta, A, A_1, K, K_1) = R_0^{Nq/2} {\cal Z}_{\rm IM} (\beta, R_1).
\label{mc}
\end{eqnarray} 
Exact expressions for the partition function of the spin-1/2 Ising model are well known for several
planar lattices \cite{onsa44,hout50,temp50,domb60,mcoy73} and accordingly, the mapping relation (\ref{mc}) 
can be rather straightforwardly employed for calculating the partition function of the model under investigation as well. Apart from this fact, it is quite evident from Eq.~(\ref{mc}) 
that the spin-1/2 Ising-Heisenberg model on a vibrating doubly decorated planar lattice becomes critical if and only if its equivalent spin-1/2 Ising model on a corresponding undecorated (rigid) lattice becomes critical as well. Critical frontiers of the spin-1/2 Ising-Heisenberg model 
on a vibrating doubly decorated planar lattice can be thus easily obtained from a comparison 
of the effective coupling (\ref{mpr2}) of the equivalent spin-1/2 Ising model with its critical value. 
For instance, the critical lines of the spin-1/2 Ising-Heisenberg model on a vibrating 
doubly decorated square lattice shown in Fig.~\ref{fig:1} can be determined from the comparison 
of the effective coupling (\ref{mpr2}) with the relevant critical point of the spin-1/2 Ising square lattice \cite{onsa44}
\begin{eqnarray}
\beta_c |R_1| = \frac{|R_1|}{k_{\rm B} T_c} = 2 \ln(1 + \sqrt{2}). 
\label{cc}
\end{eqnarray} 
Here, $T_c$ denotes the critical temperature and the absolute value $|R_1|$ reflects the 
same value of the critical temperature irrespective of whether the effective interaction 
is being ferromagnetic $R_1>0$ or antiferromagnetic $R_1<0$.

Last but not least, let us derive expressions determining the behaviour of the order parameter, 
which might be either the spontaneous magnetization for a particular case of the ferromagnetic long-range 
ordering or the spontaneous staggered magnetization for a special case of the antiferromagnetic long-range ordering. The uniform as well as staggered spontaneous magnetizations of the Ising spins $m_A$ and 
$s_A$ can readily be derived by employing exact mapping theorems developed by Barry \textit{et al}. \cite{barr88,khat90,barr91,barr95}
\begin{eqnarray}
\!\!\!\!\!\!&&\!\!\!\!\!\! m_A \!\equiv\! \frac{1}{2} \!
\left \langle \sigma_{i1}^z + \sigma_{i2}^z \right \rangle
\!=\! \frac{1}{2} \! \left \langle  \sigma_{i1}^z + \sigma_{i2}^z \right \rangle_{\rm IM} 
\! \equiv m_{\rm IM} (\beta, R_1>0)\!, \label{ma}  \\ 
\!\!\!\!\!\!&&\!\!\!\!\!\! s_A \!\equiv\! \frac{1}{2} \!
\left \langle \sigma_{i1}^z - \sigma_{i2}^z \right \rangle
\!=\! \frac{1}{2} \! \left \langle \sigma_{i1}^z - \sigma_{i2}^z \right \rangle_{\rm IM} 
\! \equiv s_{\rm IM} (\beta, R_1<0), \label{sa} 
\end{eqnarray} 
according to which the canonical ensemble average $\langle \cdots \rangle$ of any function involving 
only the Ising spins in the spin-1/2 Ising-Heisenberg model on a vibrating doubly decorating lattice directly 
equals to the canonical ensemble average $\langle \cdots \rangle_{\rm IM}$ of the same function
of the Ising spins in the equivalent spin-1/2 Ising model on a corresponding undecorated rigid lattice.
For completeness, let us quote explicit formulas for the uniform and staggered spontaneous magnetizations 
of the Ising spins in the spin-1/2 Ising-Heisenberg model on the vibrating doubly decorated square lattice, 
which are expressed in terms of the relevant exact result for the spontaneous (either uniform or staggered) 
magnetization of the spin-1/2 Ising model on the square lattice \cite{yang52,lin92}
\begin{eqnarray}
m_A &=& \frac{1}{2} \left[1 - \frac{1}{\sinh^4 \left( \beta R_1/2 \right)} \right]^{\frac{1}{8}}\!\!\!, 
\quad  (\mbox{for} \, R_1>0), \label{mae} \\ 
s_A &=& \frac{1}{2} \left[1 - \frac{1}{\sinh^4 \left( \beta |R_1|/2 \right)} \right]^{\frac{1}{8}}\!\!\!, 
\quad (\mbox{for} \, R_1<0). \label{sae}
\end{eqnarray} 

On the other hand, it is worthy of notice that the calculation of the uniform and staggered spontaneous magnetization of the Heisenberg spins $m_B$ and $s_B$ is somewhat more involved. For this purpose, one may exploit the exact generalized Callen-Suzuki spin identity \cite{call63,suzu65,saba81,saba85,balc02}
\begin{eqnarray}
\!\!&&\!\! m_B \equiv \left \langle \frac{S_{i1}^z + S_{i2}^z}{2} \right \rangle = \label{mb} \\
\!\!&&\!\! \left \langle \! \frac{1}{{\cal Z}_i} \! \iint\limits_{\!\!\!\!\!-\infty}^{\phantom{100}\infty} \!\!
  \frac{{\rm d} y_{i1} {\rm d} q_{i1}}{h} \!\! \iint\limits_{\!\!\!\!\!-\infty}^{\phantom{100}\infty} \!\! 
\frac{{\rm d} y_{i2} {\rm d} q_{i2}}{h} {\rm Tr}_{i} \!\! \left[\frac{S_{i1}^z + S_{i2}^z}{2} 
\exp(-\beta {\cal H}_i)\right] \! \right \rangle, \nonumber \\
\!\!&&\!\! s_B \equiv \left \langle \frac{S_{i1}^z - S_{i2}^z}{2} \right \rangle = \label{sb} \\
\!\!&&\!\! \left \langle \! \frac{1}{{\cal Z}_i} \! \iint\limits_{\!\!\!\!\!-\infty}^{\phantom{100}\infty} \!\!
  \frac{{\rm d} y_{i1} {\rm d} q_{i1}}{h} \!\! \iint\limits_{\!\!\!\!\!-\infty}^{\phantom{100}\infty} \!\! 
\frac{{\rm d} y_{i2} {\rm d} q_{i2}}{h} {\rm Tr}_{i} \!\! \left[\frac{S_{i1}^z - S_{i2}^z}{2} 
\exp(-\beta {\cal H}_i)\right] \! \right \rangle. \nonumber
\end{eqnarray} 
It should be nevertheless mentioned that both expressions appearing inside canonical ensemble 
average in Eqs.~(\ref{mb}) and (\ref{sb}) can be obtained following the same procedure as 
worked out previously when calculating the bond partition function ${\cal Z}_i$. The spontaneous 
magnetization of the Heisenberg spins $m_B$ can in turn be related to the spontaneous magnetization 
of the Ising spins $m_A$ through the exact formula
\begin{eqnarray}
m_B &=& m_A \frac{Q_1}{V_1},
\label{mbe}
\end{eqnarray} 
with the expression $V_1$ given by Eq.~(\ref{v1v2}) and the coefficient $Q_1$ defined as
\begin{eqnarray}
Q_1 \! = \! \exp \! \left[\frac{\beta J}{4} + \frac{\beta A_1^2}{16(K_1 + 2K)} \! \right] \!
      \sinh \! \left[ \frac{\beta J_1}{2} - \frac{\beta A A_1}{8(K_1 + 2K)} \! \right]\!\!. \nonumber \\
\label{q1}
\end{eqnarray}
After straightforward but a rather lengthly algebraic manipulation, the spontaneous staggered magnetization of the Heisenberg spins $s_B$ can also be expressed in terms of the spontaneous staggered magnetization of the Ising spins $s_A$ through the relation
\begin{eqnarray}
s_B = \frac{s_A}{V_2} \left[\frac{J_1 Q_2}{\sqrt{J_1^2 + (J \Delta)^2}} 
                            + \frac{P_2 (A Q_2 + P_1 Q_3)}{4(K_1 + 2K)} 
                      \right],
\label{sbe}
\end{eqnarray}
which contains two newly defined functions $Q_2$ and $Q_3$ 
\begin{eqnarray}
Q_2 &=& \exp \left[-\frac{\beta J}{4} + \frac{\beta P_1^2}{16(K_1 + 2 K)} \right] \nonumber \\
  &\times& \sinh \left[ \frac{\beta}{2} \sqrt{J_1^2 + (J \Delta)^2} + \frac{\beta A P_1}{8(K_1 + 2K)} \right],
\nonumber \\
Q_3 &=& \exp \left[-\frac{\beta J}{4} + \frac{\beta P_1^2}{16(K_1 + 2 K)} \right] \nonumber \\
  &\times& \cosh \left[ \frac{\beta}{2} \sqrt{J_1^2 + (J \Delta)^2} + \frac{\beta A P_1}{8(K_1 + 2K)}  \right],
\label{q2q3}
\end{eqnarray}
and the another auxiliary function $P_2$ defined as
\begin{eqnarray}
P_2 = \frac{A_1}{\sqrt{J_1^2 + (J \Delta)^2}} - \frac{J_1 P_1}{J_1^2 + (J \Delta)^2}.
\label{p2}
\end{eqnarray}

Finally, the generalized Callen-Suzuki identity \cite{call63,suzu65,saba81,saba85,balc02} will be utilized 
for a calculation of the mean displacement of both Heisenberg atoms from their equilibrium lattice positions. 
Following the same procedure as described previously by the calculation of the bond partition 
function ${\cal Z}_i$, the mean displacement of the Heisenberg atoms can be calculated with 
the help of the exact identities
\begin{eqnarray}
\left \langle y_{i1} \right \rangle &=& 
\left \langle \! \frac{1}{{\cal Z}_i} \! \iint\limits_{\!\!\!\!\!-\infty}^{\phantom{100}\infty} \!\!
  \frac{{\rm d} y_{i1} {\rm d} q_{i1}}{h} \!\! \iint\limits_{\!\!\!\!\!-\infty}^{\phantom{100}\infty} \!\! 
\frac{{\rm d} y_{i2} {\rm d} q_{i2}}{h} y_{i1} {\rm Tr}_{i} \exp(-\beta {\cal H}_i) \! \right \rangle, \nonumber \\
\left \langle y_{i2} \right \rangle &=& 
\left \langle \! \frac{1}{{\cal Z}_i} \! \iint\limits_{\!\!\!\!\!-\infty}^{\phantom{100}\infty} \!\!
  \frac{{\rm d} y_{i1} {\rm d} q_{i1}}{h} \!\! \iint\limits_{\!\!\!\!\!-\infty}^{\phantom{100}\infty} \!\! 
\frac{{\rm d} y_{i2} {\rm d} q_{i2}}{h} y_{i2} {\rm Tr}_{i} \exp(-\beta {\cal H}_i) \! \right \rangle. \nonumber
\end{eqnarray} 
It can be easily proved that the latter displacement equals identically to zero $\left \langle y_{i2} \right \rangle = 0$, 
which serves in evidence that the mean displacements of both Heisenberg atoms from the same bond are equal to each other. 
For brevity, let us merely quote the final result for the mean displacement calculated by making use of the former identity
\begin{eqnarray}
\left \langle x_{i1} \right \rangle \!&=&\! \left \langle x_{i2} \right \rangle 
= \frac{\left \langle y_{i1} \right \rangle}{\sqrt{2}}  = 
\frac{(1 - 4 \varepsilon_{\rm IM})[A (V_2 - 2Q_3) - P_1 Q_1]}{8 V_2 (K_1 + 2 K)} \nonumber \\
\!&+&\!
\frac{(1 + 4 \varepsilon_{\rm IM}) [A (V_1 - 2Q_5) - A_1 Q_1 + 2 A \Delta Q_6]}{8 V_1 (K_1 + 2 K)}.
\label{y2}
\end{eqnarray}
The newly defined functions $Q_4$--$Q_6$ entering the above formulas (\ref{y2}) are given by
\begin{eqnarray}
Q_4 \!&=&\! \exp \! \left[\! \frac{\beta J}{4} \!+\! \frac{\beta A_1^2}{16(K_1 + 2K)} \! \right] 
    \!  \cosh \! \left[\! \frac{\beta J_1}{2} \!-\! \frac{\beta A A_1}{8(K_1 + 2K)} \! \right]\!\!, 
\nonumber \\
Q_5 \!&=&\! \exp \! \left[\! -\frac{\beta J}{4} \!+\! \frac{\beta A^2 \Delta^2}{4(K_1 + 2K)} \! \right] 
\! \cosh \! \left[\! \frac{\beta J \Delta}{2} \!-\! \frac{\beta A^2 \Delta}{4(K_1 + 2K)} \! \right]\!\!, 
\nonumber \\
Q_6 \!&=&\! \exp \! \left[\! -\frac{\beta J}{4} \!+\! \frac{\beta A^2 \Delta^2}{4(K_1 + 2K)} \! \right] 
      \!  \sinh \! \left[\! \frac{\beta J \Delta}{2} \!-\! \frac{\beta A^2 \Delta}{4(K_1 + 2K)} \! \right]\!\!, \nonumber 
\label{q4q6}
\end{eqnarray}
and the quantity $\varepsilon_{\rm IM} \equiv \langle \sigma_{i1}^z \sigma_{i2}^z \rangle_{\rm IM}$
labels the pair correlation between the nearest-neighbour spins of the corresponding spin-1/2 
Ising model on undecorated rigid lattice. Hence, it follows that the foreknowledge of 
the nearest-neighbour spin correlation is needed in order to complete a calculation of 
the average displacement. For the sake of completeness, let us therefore quote also the relevant 
exact result for the nearest-neighbour pair correlation function of the spin-1/2 Ising square lattice \cite{lavi99,domb60} with the effective nearest-neighbour coupling $\beta R_1$  
\begin{eqnarray}
\varepsilon_{\rm IM} \!=\! \frac{1}{8} \coth \! \left( \! \frac{\beta R_1}{2} \! \right) \! \left \{ \!
1 + \frac{2}{\pi} \! \left[ 2 \tanh^{2} \! \left( \! \frac{\beta R_1}{2} \! \right) - 1 \right] \! 
{\cal K} (k) \! \right\},
\nonumber
\end{eqnarray}
in which the function ${\cal K} (k)$ denotes the complete elliptic integral of the first kind 
\begin{eqnarray}
{\cal K} (k) = \int_{0}^{\frac{\pi}{2}} \!\!\! \frac{{\rm d} \theta}{\sqrt{1 - k^2 \sin^2 \theta}}
\nonumber
\end{eqnarray}
with the modulus $k = 2 \sinh(\frac{\beta R_1}{2}) {\rm sech}^2 (\frac{\beta R_1}{2})$.

\section{Results and discussions}
\label{sec:result}

In this section, we will describe in detail the most interesting findings obtained for the spin-1/2 
Ising-Heisenberg model on doubly decorated planar lattices whose decorating atoms are prone to 
lattice vibrations. Before discussing the most interesting results, however, it is worthwhile 
to remark that the rigid limit ($K/J \to \infty$) of the model under investigation has been thoroughly investigated by one of the present authors in Ref.~\cite{stre02}. 
The main focus of this work is therefore to highlight the most crucial differences between the magnetic behaviour of the spin-1/2 Ising-Heisenberg model on the perfectly rigid doubly decorated lattice \cite{stre02} and respectively, the doubly decorated lattice whose decorating sites are relaxed from the condition of a perfect rigidity. Another particular goal of the present article is to explore the role of a mutual interplay 
of quantum fluctuations and lattice vibrations in determining the ground-state spin ordering, as well as, 
to examine the influence of lattice vibrations on the magnetic behaviour in a close vicinity of critical points. Note furthermore that the amplitude of lattice vibrations in the most spontaneously long-range ordered magnetic materials is usually small enough below the critical temperature and hence, our further findings will be mainly confronted with the relevant results obtained for the rigid limit of the model under consideration \cite{stre02}.  

Even though all the results derived in the foregoing section hold independently of the lattice geometry 
and also irrespective of whether exchange interactions are assumed ferromagnetic or antiferromagnetic, 
our particular attention will be hereafter restricted only to one representative example of the 
spin-1/2 Ising-Heisenberg model on a doubly decorated square lattice with both ferromagnetic exchange constants $J>0$ and $J_1>0$. 
Under this assumption, it is also reasonable to consider positive values of both magnetoelastic 
constants $A>0$ and $A_1>0$, which ensure an enhancement (reduction) of the ferromagnetic interaction 
under the contraction (elongation) of the relevant bond length. For simplicity, let us also suppose 
the equality of the interaction parameters of the same character, i.e. $J=J_1$, $A=A_1$, $K=K_1$, 
which will consequently reduce the total number of free parameters. Finally, all the interaction constants 
will be normalized hereafter with respect to a strength of the exchange interaction $J$ so that 
the ratios $k_{\rm B} T/J$, $A/J$ and $K/J$ will mark a relative magnitude of the temperature, 
the magnetoelastic constant and the spring stiffness constant, respectively. 

\begin{figure}
\begin{center}
\resizebox{0.9\columnwidth}{!}{\includegraphics{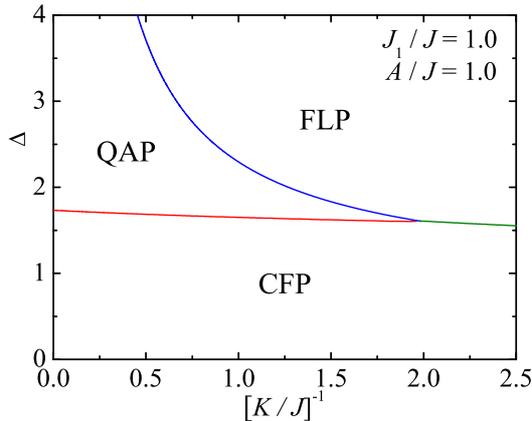}}
\end{center}
\vspace{-0.9cm}
\caption{Ground-state phase diagram in the $J/K-\Delta$ plane 
for the fixed interaction constants $J_1/J = 1.0$ and $A/J = 1.0$.}
\label{fig:2}       
\end{figure}

First, let us take a closer look at the ground-state behaviour. The typical ground-state phase 
diagram in the $J/K-\Delta$ plane is depicted in Fig.~\ref{fig:2}. Two first-order phase transition 
lines divide the ground-state phase diagram into three different regions corresponding to the classical ferromagnetic phase (CFP), the quantum antiferromagnetic phase (QAP) and the third phase with immense ditortions of lattice bonds (FLP), which is beyond the scope of present approach (see for details the 
next two paragraphs). As one would expect, the perfect ferromagnetic alignment of all Ising and Heisenberg spins can be found in CFP, which however represents the ground state only if the exchange anisotropy $\Delta$ does not exceed one of its two critical values $\Delta_{c1}$ or $\Delta_{c2}$. It is noteworthy that the highest critical value of $\Delta_{c1}^{\rm max} = \sqrt{3}$ corresponds to the rigid limit $K/J \to \infty$ and one observes a very weak gradual decline of $\Delta_{c1}$ with decreasing a relative strength of the spring stiffness constant $K/J$ that is only hardly discernible within the scale displayed in Fig.~\ref{fig:2}. On the other hand, the quite unusual QAP becomes the ground state above the critical value $\Delta_{c1}$ for strong enough spring stiffness constants. It is worthy to recall that QAP surprisingly appears in spite of ferromagnetic character of all considered exchange couplings $J = J_1>0$ 
due to a mutual competition between the easy-plane XXZ Heisenberg interaction and the easy-axis Ising 
interaction (for more details see Ref.~\cite{stre02}). The spin ordering inherent to QAP can be
characterized by a perfect antiferromagnetic (N\'eel) long-range order of the Ising spins 
from the nodal sites of a doubly decorated lattice, which is accompanied with the symmetric quantum 
superposition of two antiferromagnetic spin states of each couple of the quantum Heisenberg spins from 
the same bond of a doubly decorated lattice \cite{stre02}.\footnote{Notice that the ground state 
of the model with the antiferromagnetic Heisenberg interaction $J < 0$ is unique and it strongly resembles
the spin ordering inherent to QAP with exception that the Heisenberg spin pairs reside the antisymmetric 
quantum superposition of two antiferromagnetic spin states rather than the symmetric one.}

It should be pointed out that CFP and QAP are the only two possible ground states, which 
are separated one from each other by the critical value $\Delta_{c1}^{\rm max}=\sqrt{3}$ 
when considering the rigid limit $K/J \to \infty$ \cite{stre02}. For any finite value of the spring 
stiffness constant, however, the investigated model enters FLP whenever the exchange anisotropy 
exceeds the second critical value $\Delta_{c2}$. This latter critical value $\Delta_{c2}$ exhibits 
a rather steep decrease by shrinking the lattice rigidity, i.e. by lowering a strength of the 
spring stiffness constant $K/J$. Consequently, QAP completely vanishes from the ground-state phase diagram 
for less stiff lattices with a relatively small spring stiffness constant, where the direct 
transition from CFP to FLP takes place since the critical value $\Delta_{c2}$ 
becomes smaller than $\Delta_{c1}$ in this parameter space. 

\begin{figure}
\begin{center}
\resizebox{0.9\columnwidth}{!}{\includegraphics{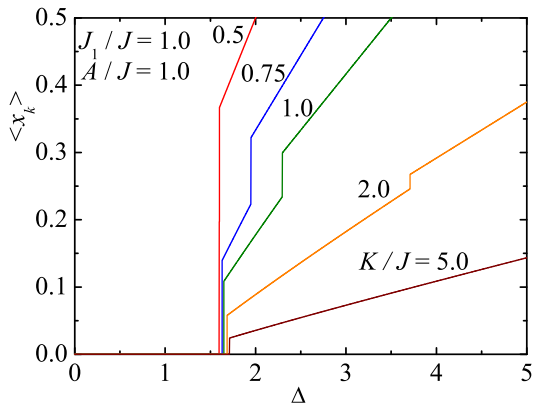}}
\end{center}
\vspace{-0.9cm}
\caption{Zero-temperature variations of the mean displacement of the Heisenberg atoms 
as a function of the exchange anisotropy $\Delta$ for $J_1/J = 1.0$, $A/J = 1.0$ and several values 
of the elastic constant $K/J$.}
\label{fig:3}       
\end{figure}  

The description of spin ordering inherent to FLP is however beyond the scope of present study, 
because one always observes in FLP a rather sizeable mean displacement of the Heisenbergs atoms
that is in contradiction with the harmonic approximation demanding small lattice deformations only.
In order to provide a deeper insight into a magnitude of those lattice deformations, Fig.~\ref{fig:3} 
shows typical zero-temperature changes of the average displacement of the Heisenberg atoms with 
the exchange anisotropy $\Delta$ for several values of the spring stiffness constant $K/J$. 
It indeed turns out that the mean displacement exhibits abrupt jumps if crossing the first-order 
phase transition lines from the ground-state phase diagram shown in Fig.~\ref{fig:2} and it 
generally increases with the exchange anisotropy when keeping the same strength of the spring stiffness constant. The formula (\ref{y2}) allows one to derive the following ground-state values for the mean displacement in the individual phases
\begin{eqnarray}
\langle x_k \rangle_{\rm CFP} &=& \frac{A-A_1}{4 (K_1 + 2K)}, \label{mdf} \\
\langle x_k \rangle_{\rm QAP} &=& \frac{2 J A \Delta^2 - J_1 A_1 - A \sqrt{J_1^2 + (J \Delta)^2}}
{4 (K_1 + 2K) \sqrt{J_1^2 + (J \Delta)^2}}, \label{mdq} \\
\langle x_k \rangle_{\rm FLP} &=& \frac{A (2 \Delta - 1)}{4 (K_1 + 2K)}. \label{mdl} 
\end{eqnarray}
It can be easily understood from Eq.~(\ref{mdf}) that the average displacement in CFP equals zero 
just if the magnetoelastic coupling constants $A$ and $A_1$ are equal, otherwise the spontaneous 
distortion of the Heisenberg and Ising bonds takes place in CFP as well. If $A>A_1$, then, the contraction 
of the Heisenberg bonds accompanied with the elongation of the Ising bonds occurs due to a larger magnetoelastic energy gain of the former bonds, while the reverse is the case if $A_1>A$. On the 
other hand, the mean displacement is always positive in QAP, which indicates shortening of the 
Heisenberg bonds at the expense of elongation of the Ising bonds on behalf of a more dominant 
magnetoelastic energy gain of the former bonds. The same holds true also for FLP, however, 
the mean displacement always gains in FLP rather high values, which are beyond the validity
of our approach. 

\begin{figure}
\begin{center}
\resizebox{0.9\columnwidth}{!}{\includegraphics{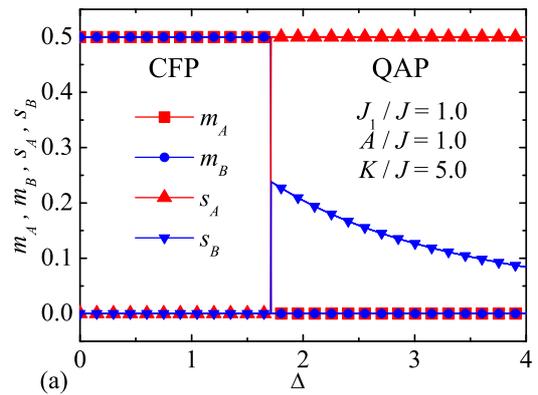}}
\resizebox{0.9\columnwidth}{!}{\includegraphics{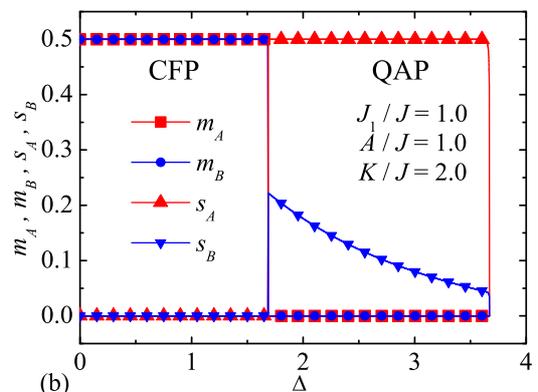}}
\end{center}
\vspace{-0.9cm}
\caption{Zero-temperature variations of the sublattice magnetizations ($m_A$, $m_B$) and the sublattice staggered magnetizations ($s_A$, $s_B$) as a function of the exchange anisotropy $\Delta$ for $J_1/J = 1.0$,  $A/J = 1.0$ and two different values of spring stiffness constant: (a) $K/J = 5.0$; (b) $K/J = 2.0$.}
\label{fig:4}       
\end{figure}

Let us close our discussion of the ground state by investigating the zero-temperature variations 
of the uniform and staggered spontaneous magnetizations in dependence on the exchange anisotropy 
$\Delta$ for two different strengths of the spring stiffness constant $K/J$ as shown in Fig.~\ref{fig:4}. 
It is noteworthy that the zero-temperature variations of the uniform and staggered spontaneous magnetizations depicted in Fig.~\ref{fig:4}(a) 
for the spring stiffness constant $K/J = 5.0$ are almost identical with the corresponding results of the perfectly rigid model ($K/J \to \infty$) 
given by
\begin{eqnarray}
\!\!\!&&\!\!\! 
\mbox{CFP:} \; [m_A, m_B, s_A, s_B] \!=\! \left[\frac{1}{2}, \frac{1}{2}, 0, 0 \right]\!\!; \label{rlf} \\
\!\!\!&&\!\!\! \mbox{QAP:} \; [m_A, m_B, s_A, s_B] 
 \!=\! \left[0, 0, \frac{1}{2}, \frac{J_1}{2 \sqrt{J_1^2 + (J \Delta)^2}} \right]\!\!. 
\label{rla}
\end{eqnarray}
While the uniform sublattice magnetizations $m_A$ and $m_B$ of both Ising as well as Heisenberg spins
achieve their saturation value in CFP and imply thus the classical character of this spontaneously 
long-range ordered phase, the staggered sublattice magnetization $s_B$ of the Heisenberg spins 
is subject to the quantum reduction of the magnetization in QAP and only the staggered sublattice magnetization $s_A$ of the Ising spins attains its saturation value. This latter finding evidently
confirms the quantum nature of QAP, which is basically influenced by pronounced local quantum fluctuations 
that also suppress the staggered sublattice magnetization $s_B$ of the Heisenberg spins. Although the zero-temperature behaviour of the uniform and staggered magnetizations remains qualitatively the same 
even for less rigid lattices, one may find an important feature which directly follows from a comparison of Fig.~\ref{fig:4}(a) and (b) displayed for a more and less rigid lattices, respectively. It turns out that the local fluctuations of lattice spacings in conjunction with the local quantum spin fluctuations tend to enhance the quantum reduction in the staggered sublattice magnetization $s_B$ of the Heisenberg spins, which originally comes from the competition between the easy-plane Heisenberg and easy-axis Ising interactions. Hence, it follows that the suppression of the staggered sublattice magnetization $s_B$ is the greater, the smaller is the spring stiffness constant $K/J$, or the higher is the magnetoelastic constant $A/J$.

\begin{figure}
\begin{center}
\resizebox{0.9\columnwidth}{!}{\includegraphics{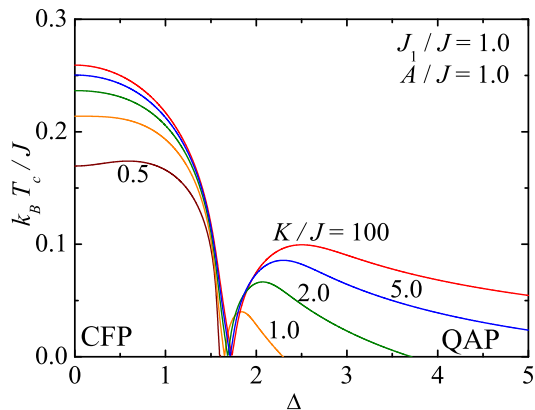}}
\end{center}
\vspace{-0.9cm}
\caption{The dependence of critical temperature on the exchange anisotropy $\Delta$ 
for $J_1/J = 1.0$, $A/J = 1.0$ and several values of the spring stiffness constant $K/J$.}
\label{fig:5}       
\end{figure}

Now, let us proceed to a discussion of the critical behaviour by investigating the finite-temperature 
phase diagram shown in Fig.~\ref{fig:5} in the form of critical temperature vs. the exchange anisotropy
dependence. It should be noted here that all the displayed critical boundaries have been obtained by solving numerically the critical condition (\ref{cc}). The finite-temperature phase diagram generally form two wings of the same critical line, which meet together at the critical value $\Delta_{c1}$ separating in the ground state CFP and QAP. The left wing of a critical line actually represents the phase boundary of CFP, whereas the right wing of a critical line represents the phase boundary of QAP. It is quite evident from Fig.~\ref{fig:5} 
that the reduction in the lattice stiffness (rigidity) generally causes the suppression in 
the critical temperature regardless of whether CFP or QAP constitutes the ground state. It should
be nevertheless stressed that there is a quite substantial difference when comparing the size 
of this reduction in CFP and QAP. For instance, the decrement in the lattice stiffness by the factor 
of 50 suppresses the critical temperature of CFP at most by 10\%, while the same decrement in the lattice stiffness may cause in QAP much more substantial diminishing of the critical temperature or even 
the complete loss of spontaneous long-range ordering  (e.g. compare the curves $K/J = 100$ and $2$ 
displayed in Fig.~\ref{fig:5}). The attenuation in the lattice rigidity may be thus 
responsible for an absence of the spontaneous ordering inherent to QAP (i.e. the right wing of 
a critical line) as it is demonstrated in Fig.~\ref{fig:5} on the particular case with $K/J = 0.5$. 
This observation would suggest that the lattice distortions basically reinforce the effect of quantum fluctuations in destroying the spontaneous long-range order of the quantum nature (such as QAP), 
while the disappearance of classical spontaneous long-range order (such as CFP) is substantially 
less affected by a presence of the lattice deformations. 

\begin{figure}
\begin{center}
\resizebox{0.9\columnwidth}{!}{\includegraphics{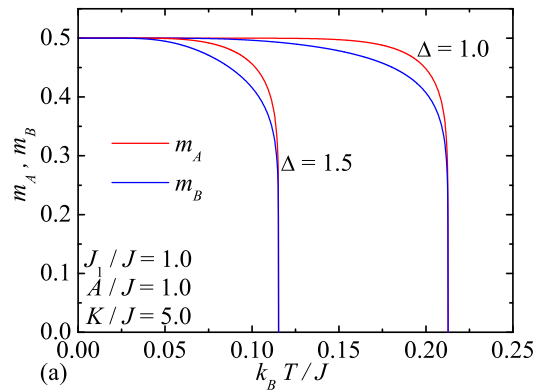}}
\resizebox{0.9\columnwidth}{!}{\includegraphics{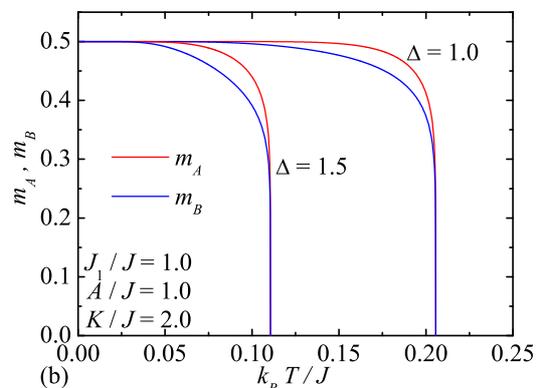}}
\end{center}
\vspace{-0.9cm}
\caption{Temperature dependences of both spontaneous sublattice magnetizations ($m_A$, $m_B$) for 
$J_1/J = 1.0$, $A/J = 1.0$, two different values of the exchange anisotropy $\Delta = 1.0$, $1.5$ 
and two different values of the spring stiffness constant: (a) $K/J = 5.0$; (b) $K/J = 2.0$.}
\label{fig:6}       
\end{figure}

Last but not least, we will turn our attention to the temperature dependences of the order parameter,
which is the uniform spontaneous magnetization for CFP and the staggered spontaneous magnetization
for QAP. Fig.~\ref{fig:6} illustrates typical thermal variations of both uniform spontaneous sublattice magnetizations for two different values of the exchange anisotropy and two different values of the spring stiffness constant. It is worthy to notice that the sublattice magnetization $m_A$ of the Ising spins is always more robust with respect to the thermal fluctuations than the sublattice magnetization $m_B$ of the Heisenberg spins even if both spontaneous sublattice magnetizations disappear in a vicinity of the critical temperature with the same critical exponent $\beta_e = 1/8$ from the standard Ising universality class. It can be also clearly seen from the comparison of Fig.~\ref{fig:6}(a) and (b) that the reduction in the lattice stiffness does not change basically the temperature dependences of the spontaneous magnetizations, but it only causes a small gradual shift of the critical temperature towards lower values.

\begin{figure}
\begin{center}
\resizebox{0.9\columnwidth}{!}{\includegraphics{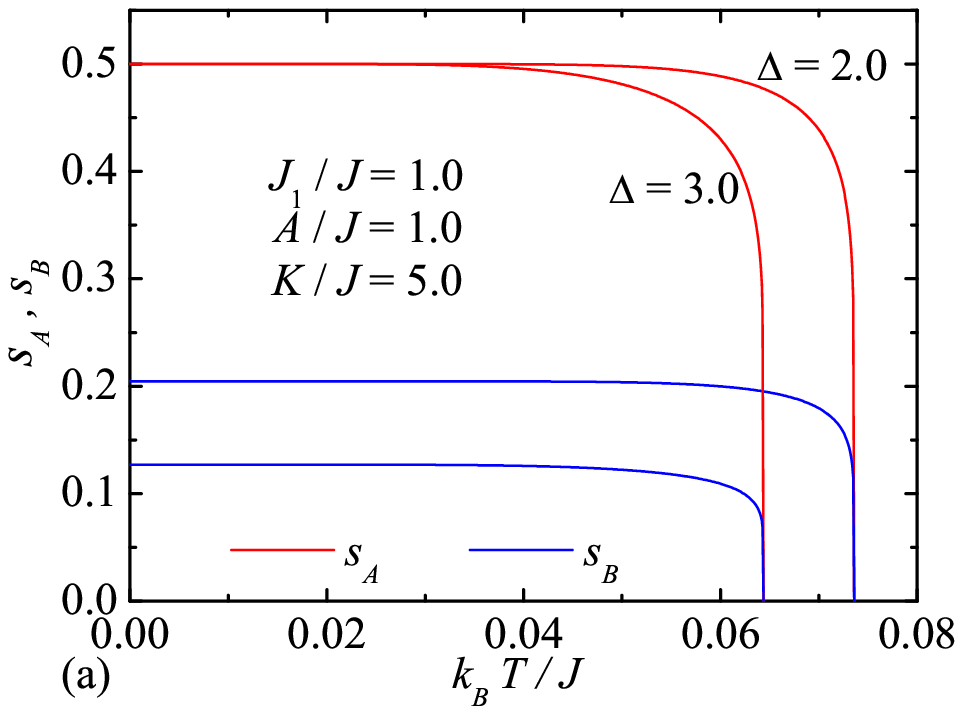}}
\resizebox{0.9\columnwidth}{!}{\includegraphics{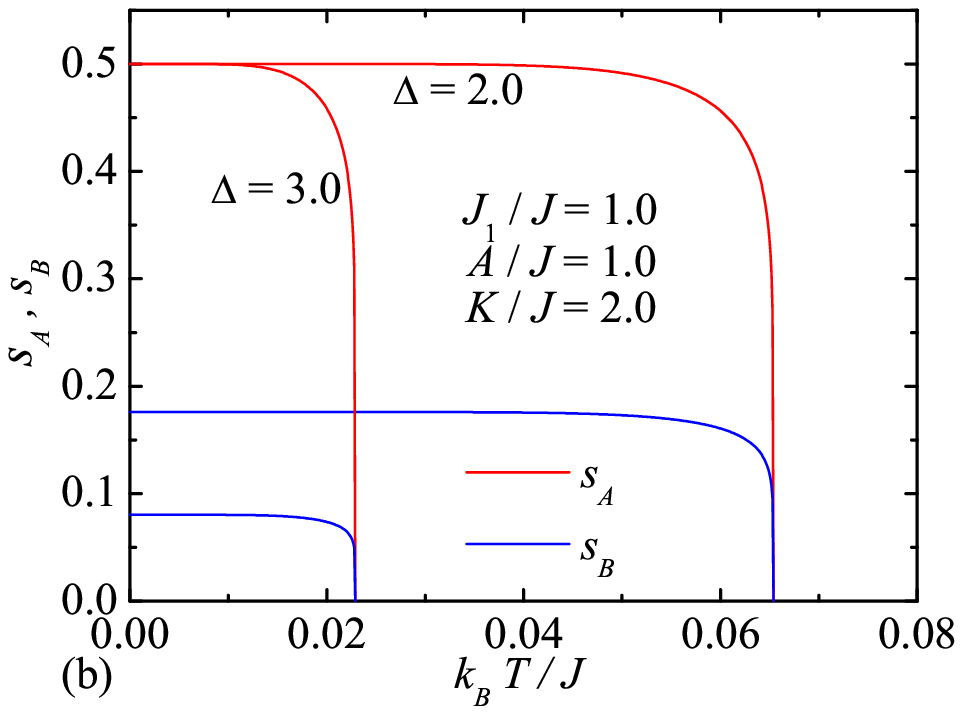}}
\end{center}
\vspace{-0.9cm}
\caption{Temperature dependences of both staggered sublattice magnetizations ($s_A$, $s_B$) 
for $J_1/J = 1.0$, $A/J = 1.0$, two different values of the exchange anisotropy $\Delta = 2.0$, $3.0$ 
and two different values of the spring stiffness constant: (a) $K/J = 5.0$; (b) $K/J = 2.0$.}
\label{fig:7}       
\end{figure}

The relevant temperature behaviour of both staggered spontaneous sublattice magnetizations 
in QAP is illustrated in Fig.~\ref{fig:7}. While the critical exponent determining the 
variation of both staggered sublattice magnetizations in a close vicinity of the critical temperature 
remains the same $\beta_{e} = 1/8$, the drop in the lattice rigidity has much more obvious 
effect upon the thermal dependences of the staggered sublattice magnetizations. 
It can be indeed readily understood from the comparison of Fig.~\ref{fig:7}(a) and (b)
that the staggered sublattice magnetization $s_B$ starts from much lower initial value when 
decreasing the lattice stiffness, because it becomes subject of a stronger quantum reduction 
of the magnetization due to the interplay between the local quantum spin fluctuations and the 
local fluctuations of lattice spacings. In addition, one also observes a more substantial reduction 
in the critical temperature by lowering the spring stiffness constant, which comes from a mutual 
interplay between the temperature-induced spin fluctuations, the quantum spin fluctuations as well as 
the distortion of lattice bonds.

\section{Conclusion}
\label{sec:conc}

The present article deals with the magnetoelastic properties of the spin-1/2 Ising-Heisenberg model
defined on doubly decorated planar lattices whose decorating atoms may oscillate around their 
equilibrium lattice positions after taking into account the elastic energy penalty of each distorted
bond within the harmonic approximation. It has been shown that after tracing out the 
spin degrees of freedom of the Heisenberg spins and integrating out the lattice degrees of freedom 
it is possible to establish a simple mapping equivalence between the investigated model system 
and its corresponding spin-1/2 Ising model on undecorated rigid lattice using the generalized 
decoration-iteration transformation \cite{fish59,roja09,stre10,lap}. 
With the help of this mapping method, we have examined the ground-state and finite-temperature phase diagrams, 
the magnitude of the mean displacement, as well as, the temperature dependences of the order parameters.

To shed light on the influence of the lattice vibrations, our particular attention has been devoted 
to the most significant differences between the magnetic behaviour of the spin-1/2 Ising-Heisenberg 
model on the perfectly rigid doubly decorated lattice \cite{stre02} and respectively, the doubly 
decorated lattice whose decorating atoms are relaxed from the condition of a perfect rigidity. 
It has been demonstrated that the mutual interplay between the quantum spin fluctuations and 
the lattice deformations enhances typical quantum features like the quantum reduction of the magnetization 
in the ground state of quantum phases such as QAP, while it does not affect the respective ground-state behaviour 
of classical phases such as CFP by no means. As far as the magnetic behaviour at finite temperatures is concerned, 
it also turns out the fluctuations in lattice spacings act more efficiently in conjunction with thermal 
and quantum spin fluctuations in destroying a spontaneous long-range order of the quantum phases, whereas the 
influence of the lattice distortions in promoting thermal and quantum fluctuations by disturbing a spontaneous 
long-range order of the classical phases is much less efficient and conspicuous.

Finally, let us briefly mention that theoretical treatment of lattice vibrations within the framework of classical 
harmonic oscillators rather than quantum harmonic oscillators represents perhaps the most important drawback of the proposed model. 
It should be nevertheless mentioned that the approach presented in this paper can be rather straightforwardly generalized 
to the spin-1/2 Ising-Heisenberg model on doubly decorated planar lattices whose lattice vibrations will be 
treated as quantum harmonic oscillators as well. Our preliminary study of this latter more
sophisticated model reveal the completely same results for the ground-state and finite-temperature
phase diagrams, as well as, the relevant behaviour of the order parameter due to the same form 
of the magnetic part of the bond partition function \cite{roja11}. The most fundamental differences 
between the Ising-Heisenberg model with the lattice vibrations described at the classical and quantum level, 
respectively, thus appears in a respective behaviour of basic thermodynamic quantities (such as internal energy, 
entropy or specific heat) that are also affected by the elastic part of the bond partition function. For instance, 
the former model apparently exhibits artificial temperature dependences of the specific heat owing to the anomalous 
lattice contribution of the heat capacity, but we have not dwell on investigating this aspect of our model in the present paper. 

\begin{acknowledgement}
The financial support of this work provided by the research foundation FAPEMIG 
under the grant number BPV-00088-10 is gratefully acknowledged.
\end{acknowledgement}

\end{document}